\documentclass[lettersize,journal]{IEEEtran}
\usepackage{amsmath,amsfonts}
\usepackage{algorithmic}
\usepackage{algorithm}
\usepackage{array}
\usepackage[caption=false,font=normalsize,labelfont=sf,textfont=sf]{subfig}
\usepackage{textcomp}
\usepackage{stfloats}
\usepackage{url}
\usepackage{verbatim}
\usepackage{graphicx}
\usepackage{cite} 

\begin{document}

\title{Precise Representation Model of SAR Saturated Interference: 
Mechanism and Verification
} 

\author{Lunhao~Duan,
Xingyu~Lu,
Yushuang~Liu,
Jianchao~Yang,
Hong~Gu

\thanks{Lunhao Duan, Xingyu Lu, Yushuang Liu, Jianchao Yang and Hong Gu
are with the Department of Electronic and Optical Engineering, Nanjing
University of Science and Technology, Nanjing 210094, China (e-mail:
{nust\_luxyhenry@163.com}).}}

\maketitle

\begin{abstract}
Synthetic Aperture Radar (SAR) is highly susceptible to Radio Frequency Interference (RFI). Due to the performance limitations of components such as gain controllers and analog-to-digital converters in SAR receivers, high-power interference can easily cause saturation of the SAR receiver, resulting in nonlinear distortion of the interfered echoes, which are distorted in both the time domain and frequency domain. Some scholars have analyzed the impact of SAR receiver saturation on target echoes through simulations. However, the saturation function has non-smooth characteristics, making it difficult to conduct accurate analysis using traditional analytical methods. Current related studies have approximated and analyzed the saturation function based on the hyperbolic tangent function, but there are approximation errors. Therefore, this paper proposes a saturation interference analysis model based on Bessel functions, and verifies the accuracy of the proposed saturation interference analysis model by simulating and comparing it with the traditional saturation model based on smooth function approximation. This model can provide certain guidance for further work such as saturation interference suppression.
\end{abstract}


\section{Introduction}
\IEEEPARstart{S}{ynthetic} aperture radar (SAR) has become a powerful remote sensing tool due to its all-weather, all-time imaging capability and high-resolution image quality. However, with the rapid proliferation of ground-based radar, communication equipment, and spaceborne SAR systems, SAR is highly susceptible to Radio Frequency Interference (RFI) during operation\cite{shu2024sar}.

Accordingly, scholars have conducted in-depth research from the perspectives of parameterization, non-parameterization, and semi-parameterization, and most methods have certain effects on RFI in SAR.
Parameterization methods aim to suppress interference by estimating interference parameters, reconstructing, and extracting interference. Zhou et al.\cite{zhou2007time} proposed a new method for narrowband interference suppression based on an improved least mean square algorithm. Liu et al.\cite{liu2013time} put forward an RFI suppression method combining the iterative adaptive approach and orthogonal subspace projection.

Non-parameterization methods intend to filter out interference by utilizing the characteristic differences between interference and target echoes. Tao et al. \cite{tao2015wideband} segmented broadband interference and removed the interference in each time segment using frequency-domain filtering. Yang et al.\cite{yang2021bsf} performed interference suppression by constructing a block subspace filter in SAR complex images.

Semi-parameterization methods convert complex signal separation problems into hyperparameter optimization problems. The main idea of this method is to suppress interference by leveraging the sparsity of interfered SAR echoes. Su et al.\cite{Su9915479} decomposed the time-frequency pulse matrix into useful signals and interference based on the low-rank and sparse characteristics of interfered echoes, and finally subtracted the interference from the echoes to achieve interference suppression. Huang et al. \cite{8930046,huang2022efficient} proposed a series of interference suppression algorithms based on low-rank recovery theory and alternating projection. Li et al. \cite{li2022pulse} located RFI through eigenvalue decomposition and short-time Fourier transform, and then suppressed the interference using a time-domain notch filter.

Most of these methods exhibit certain effectiveness for RFI in SAR. Nevertheless, these approaches are only applicable to scenarios where RFI is unsaturated, i.e., cases where RFI and target echoes are linearly superimposed, and they are not suitable for nonlinear superposition scenarios. Herein, the nonlinear superposition primarily stems from the saturation of the SAR receiver. Due to performance limitations of components such as gain controllers and analog-to-digital converters (ADCs) in the SAR receiver, high-power interference can easily cause receiver saturation, leading to nonlinear distortion of the interfered echoes, which manifests as distortions in both time and frequency domains.

Saturated interference in SAR is not uncommon. Cases of SAR suffering from saturated interference have been reported in the LT-1 spaceborne SAR\cite{cai2023first} and a certain type of Chinese airborne SAR \cite{shen2023research}, respectively. Since the power of the interference directly reaching the SAR receiver undergoes only one-way attenuation, it is much stronger than the power of the echo reflected by the target. Consequently, it can easily cause saturation of the SAR receiving channel, thereby inducing nonlinear clipping distortion in the receiving channel. This distortion leads to the degradation of time-domain and frequency-domain characteristics of the echo signal: in the time domain, the signal amplitude is clipped, resulting in amplitude and phase distortion; in the frequency domain, the nonlinear nature of saturation distortion introduces spectral spurs. If the particularities of saturated interference are ignored and existing interference analysis and suppression methods are still employed, it will lead to mismatches between the signal model and the suppression method, resulting in poor interference suppression performance.

Regarding the saturation problem in SAR received signals, references\cite{li2011effects,li2013effect} simulated and analyzed the impact of SAR receiver saturation distortion on echo signals and SAR imaging, but did not investigate the analytical characteristics or time-frequency behavior of saturated echoes when interference is present in the echoes. Reference\cite{xiao2019reconstruction} elaborated on the causes of receiver saturation and proposed a recovery method for saturated SAR echoes, but the authors also did not consider the presence of interference. In the context of single-bit sampling, \cite{franceschetti2002theory,franceschetti1991processing,zhao2019one} analyzed the components of echoes after single-bit sampling and proposed corresponding single-bit sampling models. However, single-bit sampling is a special case of analog-to-digital converter (ADC) sampling, which is effective for analyzing saturated interference suppression in single-bit sampling techniques. Nevertheless, this model is no longer applicable to receivers with other sampling bit depths, and thus only has limited reference value. Additionally, the authors used a sign function to represent the single-bit sampling model. However, the sign function is generally a binary function that is discontinuous and has a jump process, whereas saturation is a continuous process without abrupt jumps\cite{wu2011redesign}, leading to certain differences between the two. 

Although the aforementioned studies have provided theoretical analyses of SAR echo saturation, none of them have considered the presence of interference in the echoes. In reality, the existence of interference signals introduces more complex issues such as harmonic distortion and nonlinear cross-interference, making it more difficult to model and analyze SAR received signals subjected to saturation interference.
Reference \cite{duan2025saturated} constructed an approximate analysis model of SAR saturation interference response based on the $\mathrm{tanh}$ function, and proposed a suppression method for high-order harmonics of interference based on this model. However, approximate processing was adopted in deriving this model, with high-order exponential terms neglected, resulting in incomplete accuracy of the amplitude information of each harmonic component in the model.
In view of the above factors, to achieve more accurate results and provide guidance for further saturation interference suppression, this paper establishes an accurate SAR saturation interference analysis model based on Bessel functions.

\section{Precise Signal Model for SAR Saturated Interference}
\subsection{Analysis of Saturation Mechanisms and Development of a Saturated Interference Model}
SAR receiver typically consists of components such as an antenna front-end, gain controller, filter, and ADC. A simplified architecture diagram of the SAR receiver is presented in Fig. 1.
\begin{figure}[htbp]
    \centering
    \includegraphics[width=0.45\textwidth]{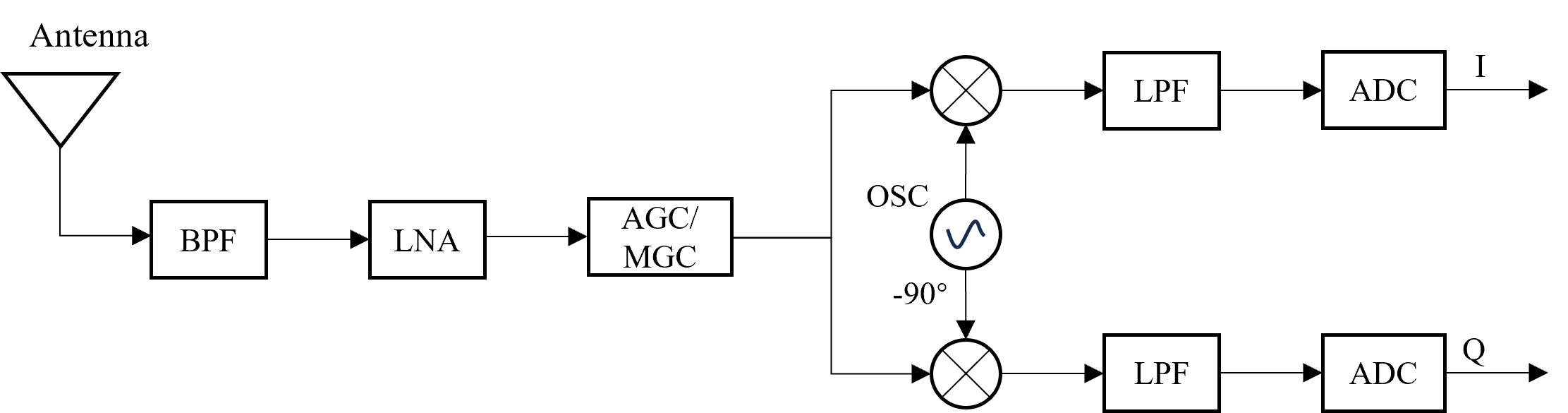} 
    \caption{Simplified Architecture Diagram of Synthetic Aperture Radar Receiver} 
    \label{fig:sar_interference} 
\end{figure}
 
Among these components, gain controllers can be categorized into automatic gain controllers (AGCs) and manual gain controllers (MGCs). Their primary function is to regulate the intensity of received echoes, preventing their amplitude from exceeding the dynamic range of the ADC. When the amplitude of the input signal lies within the ADC's dynamic range, the ADC typically operates in a linear state. However, once the input signal amplitude exceeds this dynamic range, the ADC enters a nonlinear operating state, resulting in signal saturation. In practice, gain controllers usually adjust the gain based on echoes following a zero-mean complex Gaussian distribution. To achieve high sensitivity, they generally overlook scenarios where strong scatterers or intense interference exist in the echoes\cite{xiao2019reconstruction}. If strong interference is present in the received echoes, the gain controller fails to effectively regulate the intensity of the interfered echoes, causing the input signal amplitude to the ADC to become excessively large, and thus leading to signal saturation.

Assuming the dynamic range of the ADC is$[-s_a,s_a]$, the saturation process can be described by the following mathematical model:
\begin{equation}s_{out}=sat(x)\end{equation}
 where
\begin{equation}
sat(x) = 
\begin{cases} 
s_a & x > s_a \\
x & |x| < s_a \\
-s_a & x < -s_a 
\end{cases}
\end{equation}
where $sat(\cdot)$ denotes the saturation function, $x$ represents the input analog signal, and $s_{out}$ denotes the output digital signal. In this paper, the ratio of the maximum amplitude of the saturated output signal to the maximum amplitude of the unsaturated signal is defined as the saturation coefficient, which is less than 1.

\subsection{Mathematical Model of Interfered Echoes Under Saturation Conditions}
After a Synthetic Aperture Radar (SAR) transmits a signal, the signal returning to the receiver antenna is generally regarded as a mixed signal consisting of target echo signals, interference, and random noise. Compared with the first two components, the noise power is relatively low, so its influence can be neglected. In addition, since frequency mixing and filtering are linear processes, the input signal to the ADC can be approximately expressed as follows:
\begin{equation}
s_{in}(t) = e(t) + i(t) = \alpha \exp(\mathrm{j}\varphi) + b\exp(\mathrm{j}\xi)
\end{equation}
where $t$ denotes the range fast time, $s_{in}(t)$ represents the SAR received signal, $e(t)$ and $i(t)$ denote the target echo and interference signal respectively, $\varphi$ and $\xi$ are the phases of the target echo and interference, respectively, while $a$ and $b$ are their amplitudes.

Assuming that the ADC saturation is caused by the presence of high-power interference, the derivation of the harmonic component model for saturation interference is presented below. Since the $\mathrm{tanh}$ function is formally similar to the saturation function (2) and is a continuously differentiable smooth function, it can be expanded directly using the following Taylor series. 

\begin{equation}
\begin{split}
\tanh\left(x\right) &= \sum_{n=1}^{\infty}\frac{2^{2n}\left(2^{2n}-1\right)B_{2n}x^{2n-1}}{\left(2n\right)!}\\
&= x - \frac{x^{3}}{3} + \frac{2x^{5}}{15} - \frac{17x^{7}}{315} + \cdots\cdots\quad|x| < \frac{\pi}{2}
\end{split}
\end{equation}

Therefore, reference \cite{duan2025saturated} approximates the saturation function (2) as Equation (5).
\begin{equation}
g(s_{in}) = s_a\cdot\mathrm{tanh}(\frac{s_{in}}{s_a})
\end{equation}

Substituting Equations (3) and (4) into Equation (5) yields the approximate saturation analysis model.

From the above analysis and derivation, it can be seen that the saturation model based on the $\mathrm{tanh}$ function involves two approximations: first, directly approximating the saturation function as the tanh function, which are not completely equivalent and thus introduce errors; second, the authors neglected the high-order terms of the Taylor series in Equation (5) during subsequent processing, which also leads to errors. These approximations result in accuracy errors in the tanh-based saturation analysis model, making it unable to accurately describe the saturation model. Therefore, this paper proposes an accurate saturation analysis model, which is introduced below.

According to references \cite{franceschetti2002theory,franceschetti1991processing,zhao2019one}, we can convert the saturation function (2) into the following integral form:

\begin{equation}sat(x)=\frac{-j}{\pi}\int_{-\infty}^{\infty}(\frac{\sin(s_{a}w)}{w^{2}})e^{jwx}dw\end{equation}

Substituting the ADC input signal in Equation (3) into Equation (6) to simulate saturation, then we can obtain the saturated output signal, as shown in Equation (7).
\begin{equation}
\begin{split}
s_{\text{out}} &= \text{sat}(s_{\text{in}}) = \frac{-\mathrm{j}}{\pi}\int_{-\infty}^{\infty}\frac{\sin\left(s_a w\right)}{w^2}\mathrm{e}^{\mathrm{j}w(e + t)dw}dw\\
&= \frac{-\mathrm{j}}{\pi}\int_{-\infty}^{\infty}A_1\,\exp\bigl[\mathrm{j}w(a\cos\varphi + b\cos\xi)\bigr]dw \\
&\quad + \frac{1}{\pi}\int_{-\infty}^{\infty}A_1\,\exp\bigl[\mathrm{j}w(a\sin\varphi + b\sin\xi)\bigr]dw\\
&= \frac{-\mathrm{j}}{\pi}\int_{-\infty}^{\infty}A_1\,\exp(\mathrm{j}wa\cos\varphi)\exp(\mathrm{j}wb\cos\xi)dw \\
&\quad + \frac{1}{\pi}\int_{-\infty}^{\infty}A_1\,\exp(\mathrm{j}wa\sin\varphi)\exp(\mathrm{j}wb\sin\xi)dw
\end{split}
\end{equation}
where $A_1=\frac{\mathrm{sin}(s_aw)}{w^2}$.

According to \cite{zhao2019one},
\begin{equation}
\mathrm{exp}(\mathrm{j}wa\mathrm{cos}\varphi) = \sum_{m=0}^{\infty}\alpha_m\mathrm{j}^mJ_m(aw)\mathrm{cos}(m\varphi)
\end{equation}
\begin{equation}
\mathrm{exp}(\mathrm{j}wa\mathrm{sin}\varphi) = \sum_{m=0}^{\infty}\alpha_m\mathrm{-j}^mJ_m(aw)\mathrm{cos}[m(\varphi+\frac{\pi}{2})]
\end{equation}
where $J_m(x)$ denotes the $m$-th order Bessel function. $\alpha_0=1$, and $\alpha_m=2$ when $m\geq1$. Substituting Equations (8) and (9) into Equation (7) and rearranging, the final saturated interference analysis model is obtained as shown in Equation (10).

\begin{figure}[hb!]
	\centering
	\subfloat[]
	{\includegraphics[width=0.23\textwidth]{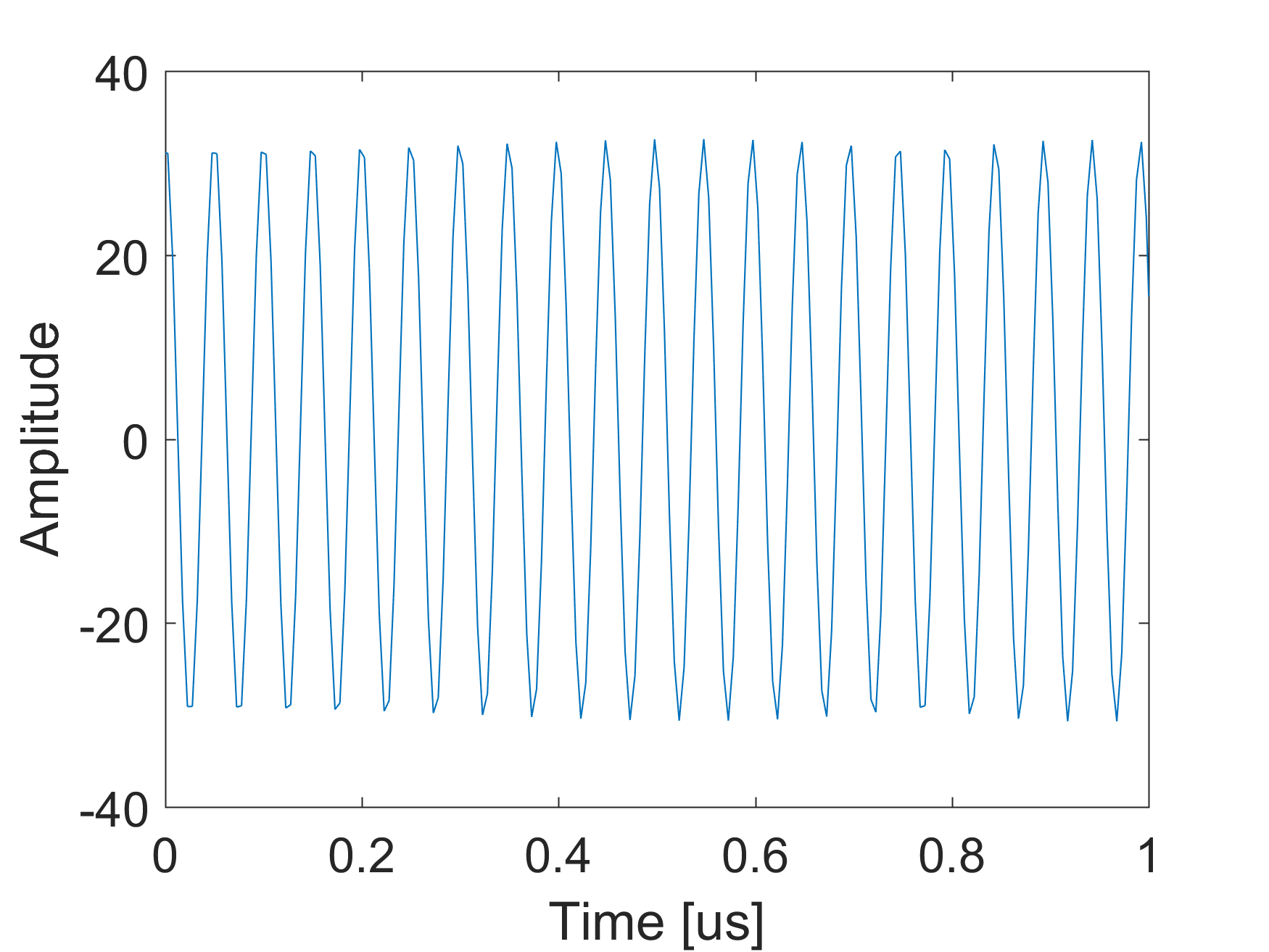}\label{figures/echo_before_sat.png}}
	\subfloat[]
	{\includegraphics[width=0.23\textwidth]{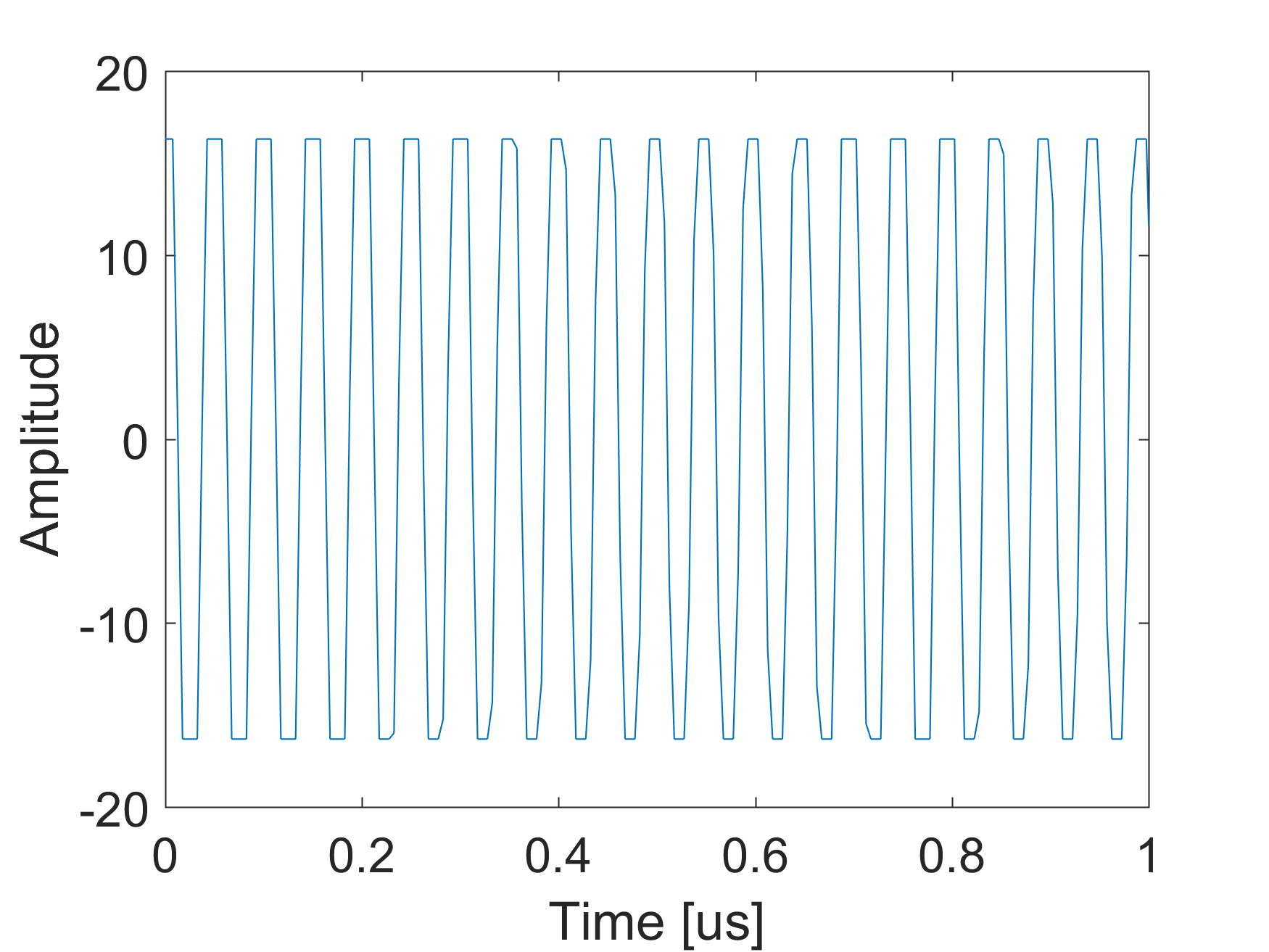}\label{figures/echo_after_sat.png}}
	\caption{Time-Domain Waveforms Before and After Saturation(a) Unsaturated Time Domain(b) Saturated Time Domain}
\end{figure}

\begin{figure*}
\begin{equation}
\begin{split}
s_{\text{out}} &= \frac{-\mathrm{j}}{\pi} \int_{-\infty}^{\infty} A_1 \sum_{m=0}^{\infty} \sum_{n=0}^{\infty} \alpha_m \alpha_n \, \mathrm{j}^{m+n} J_m(a w) J_n(b w) \cos(m\varphi) \cos(n\xi) \, dw \\
&\quad + \frac{1}{\pi} \int_{-\infty}^{\infty} A_1 \sum_{m=0}^{\infty} \sum_{n=0}^{\infty} \alpha_m \alpha_n \, (-\mathrm{j})^{m+n} J_m(a w) J_n(b w) \cos\left[m\left(\varphi + \frac{\pi}{2}\right)\right] \cos\left[n\left(\xi + \frac{\pi}{2}\right)\right] \, dw \\
&= \frac{-1}{\pi} \sum_{m=0}^{\infty} \sum_{n=0}^{\infty} \alpha_m \alpha_n A_2 \left\{ \mathrm{j}^{m+n+1} \cos(m\varphi) \cos(n\xi) - (-\mathrm{j})^{m+n} \cos\left[m\left(\varphi + \frac{\pi}{2}\right)\right] \cos\left[n\left(\xi + \frac{\pi}{2}\right)\right] \right\} \\
&= \frac{-1}{2\pi} \sum_{m=0}^{\infty} \sum_{n=0}^{\infty} \alpha_m \alpha_n A_2 \, (-1)^{\frac{m+n+1}{2}} \left\{ \exp\left[ (-1)^{\frac{m+n+3}{2}} \mathrm{j}(m\varphi + n\xi) \right] + \exp\left[ (-1)^{\frac{m-n+3}{2}} \mathrm{j}(m\varphi - n\xi) \right] \right\}
\end{split}
\end{equation} 
\end{figure*}
where 
\begin{equation}
\begin{split}
A_2&=\int_{-\infty}^{\infty}A_1J_m(aw)J_n(bw)dw\\
&=\int_{-\infty}^{\infty}\frac{\sin(s_aw)}{w^2}J_m(aw)J_n(bw)dw
\end{split}
\end{equation}

It can be seen that the constituent components of the saturated signal consist of several harmonics. When $m=0$, the phase of the harmonic only contains the phase $\xi$ of the original interference and is irrelevant to the phase $\varphi$ of the original target echo, which is called the interference harmonic component. Furthermore, the harmonic component when $m=0,n=1$ is referred to as the interference fundamental wave, and the harmonic components when $m=0,n\neq1$ are called interference higher-order harmonics. Similarly, the harmonic component when $m=1,n=0$ is called the target fundamental harmonic, and the harmonic components when $m\neq1,n=0$ are referred to as target higher-order harmonics. When $m\neq0,n\neq0$, the phase of the harmonic is related to both the phase $\xi$  of the original interference and the phase $\varphi$ of the original target echo, which is called the cross-term harmonic between the interference and the target echo.

\section{Verification of the Saturated Model and Analysis of Coupling Characteristics}
The precise saturated model of interfered saturated echoes has been established in the previous section. To further explore the variation laws of each component in the saturated output model, this section verifies the accuracy of the saturated output model through simulations and conducts a detailed analysis of each harmonic component in the model.
 
This subsection verifies the accuracy of the saturated interference model (10) through simulations. Two linear frequency modulation (LFM) signals are set as the target echo and interference signal, respectively, with a pulse width of 30$\mu$s for both. The target echo has a bandwidth of 5 MHz and a center frequency of 0 Hz, while the interference signal has a bandwidth of 10 MHz and a center frequency of 20 MHz. The Interference-to-Signal Ratio (ISR) is set to 30 dB, and the saturation coefficient is 0.5. Detailed parameters are as follows: $a=1,b=31.62,s_a=16.31$. The simulation results are shown in Fig. 2 and Fig. 3.

\begin{figure*}[hb]
	\centering
 
	\subfloat[]{\includegraphics[width=2in]{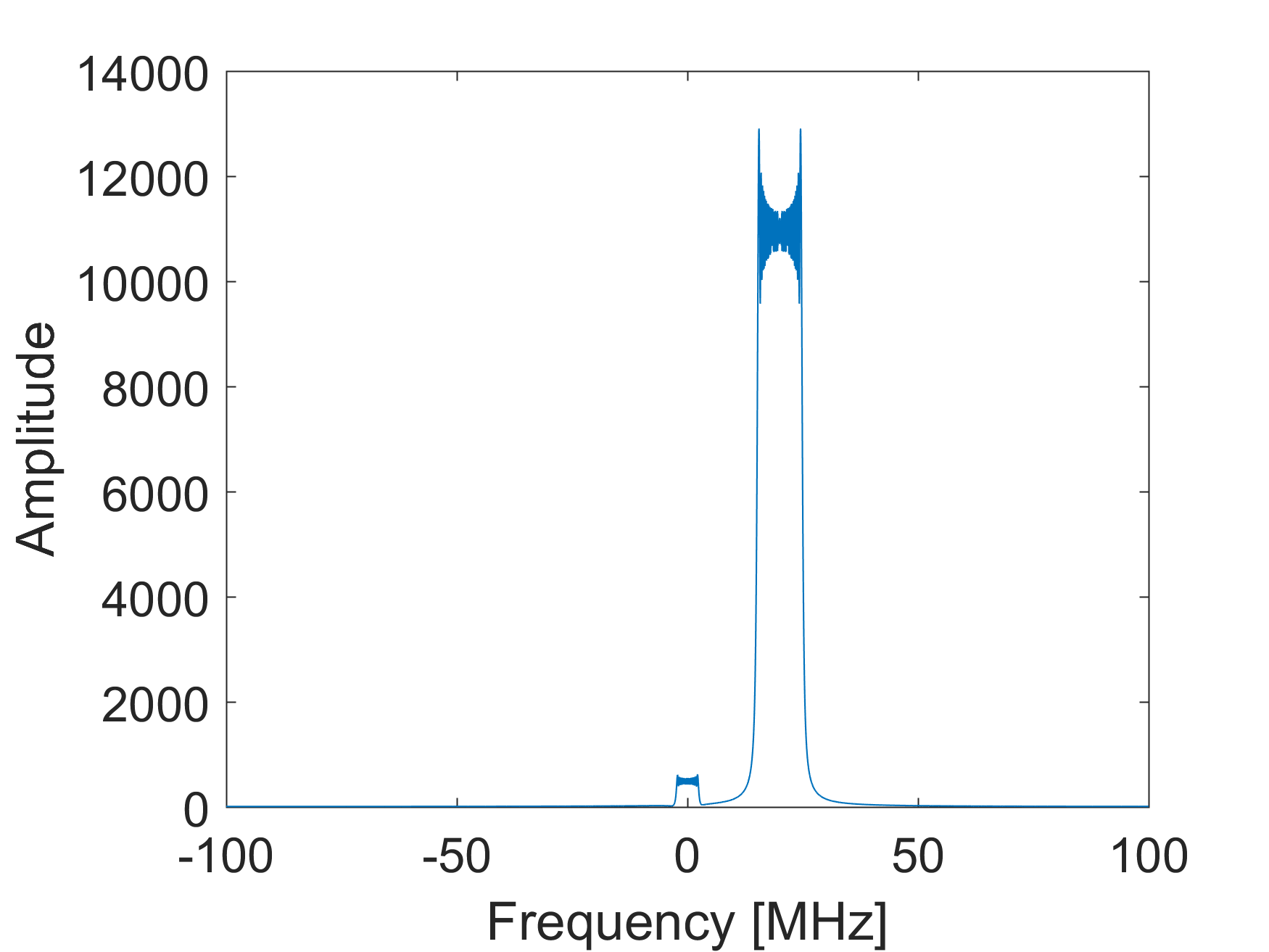}%
		\label{1}}\hspace{0.01in} 
	\subfloat[]{\includegraphics[width=2in]{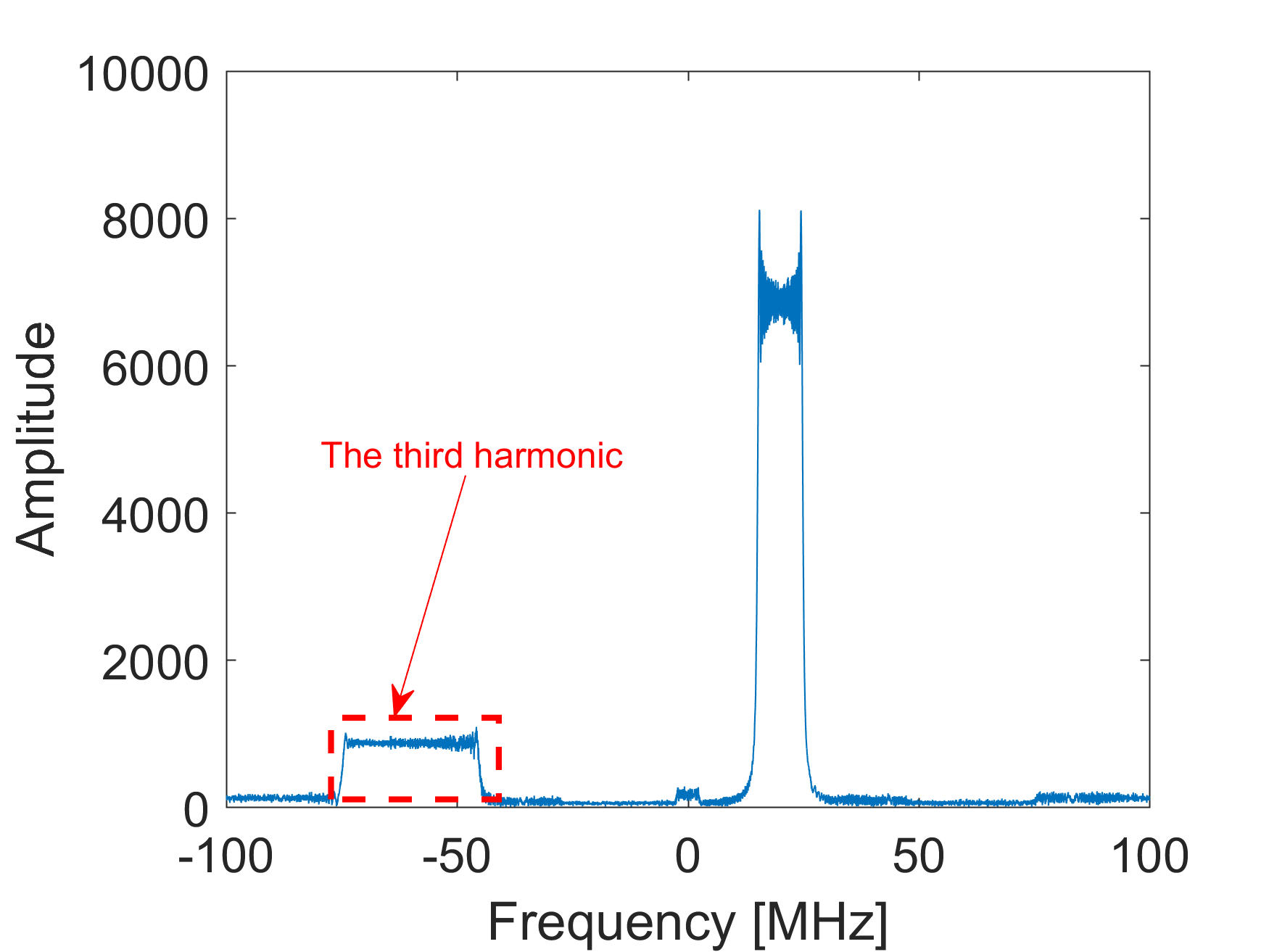}%
		\label{fig_second_case66}}\hspace{0.01in} 
	\subfloat[]{\includegraphics[width=2in]{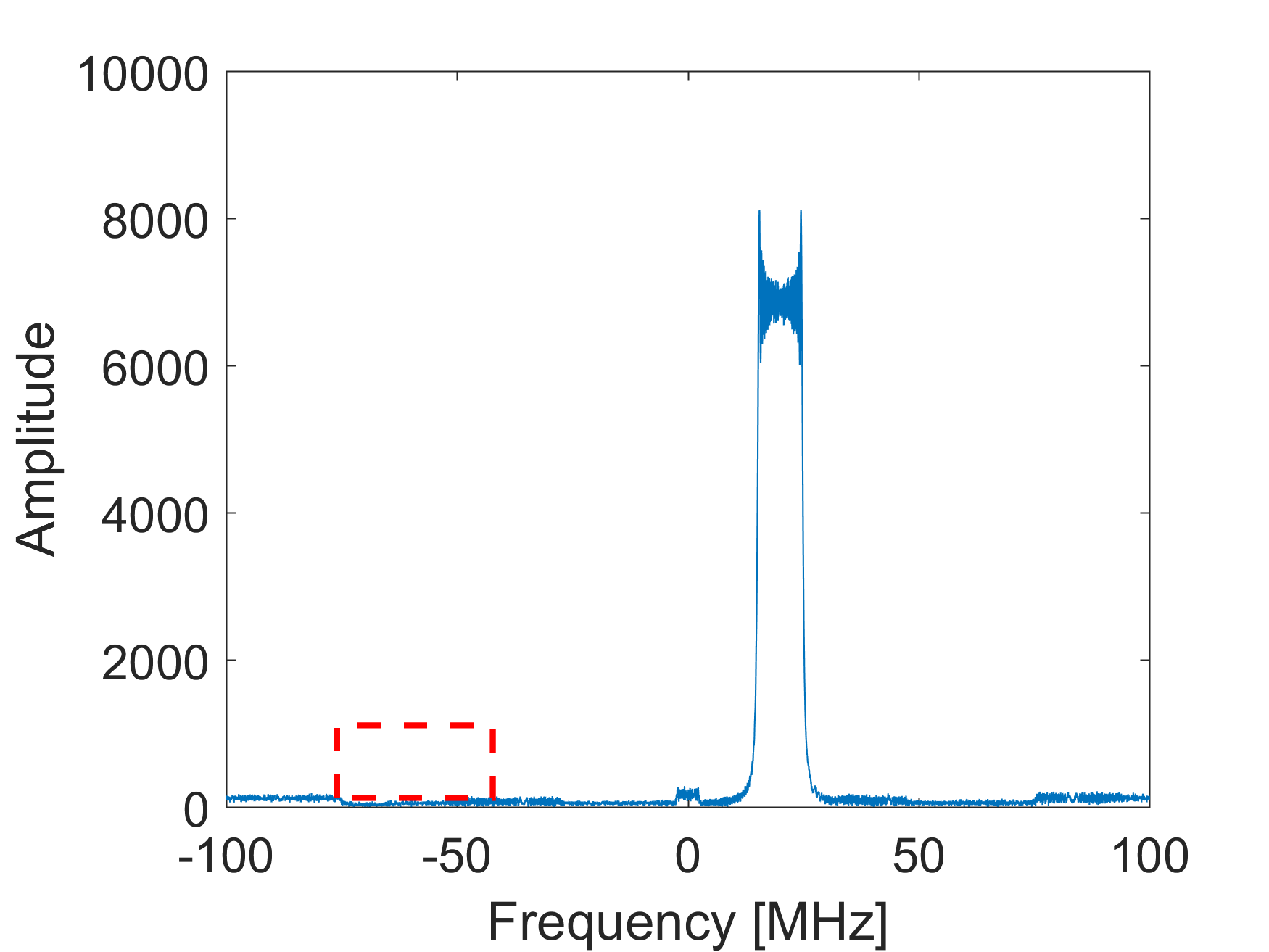}%
		\label{fig_first_case7}}

	\subfloat[]{\includegraphics[width=2in]{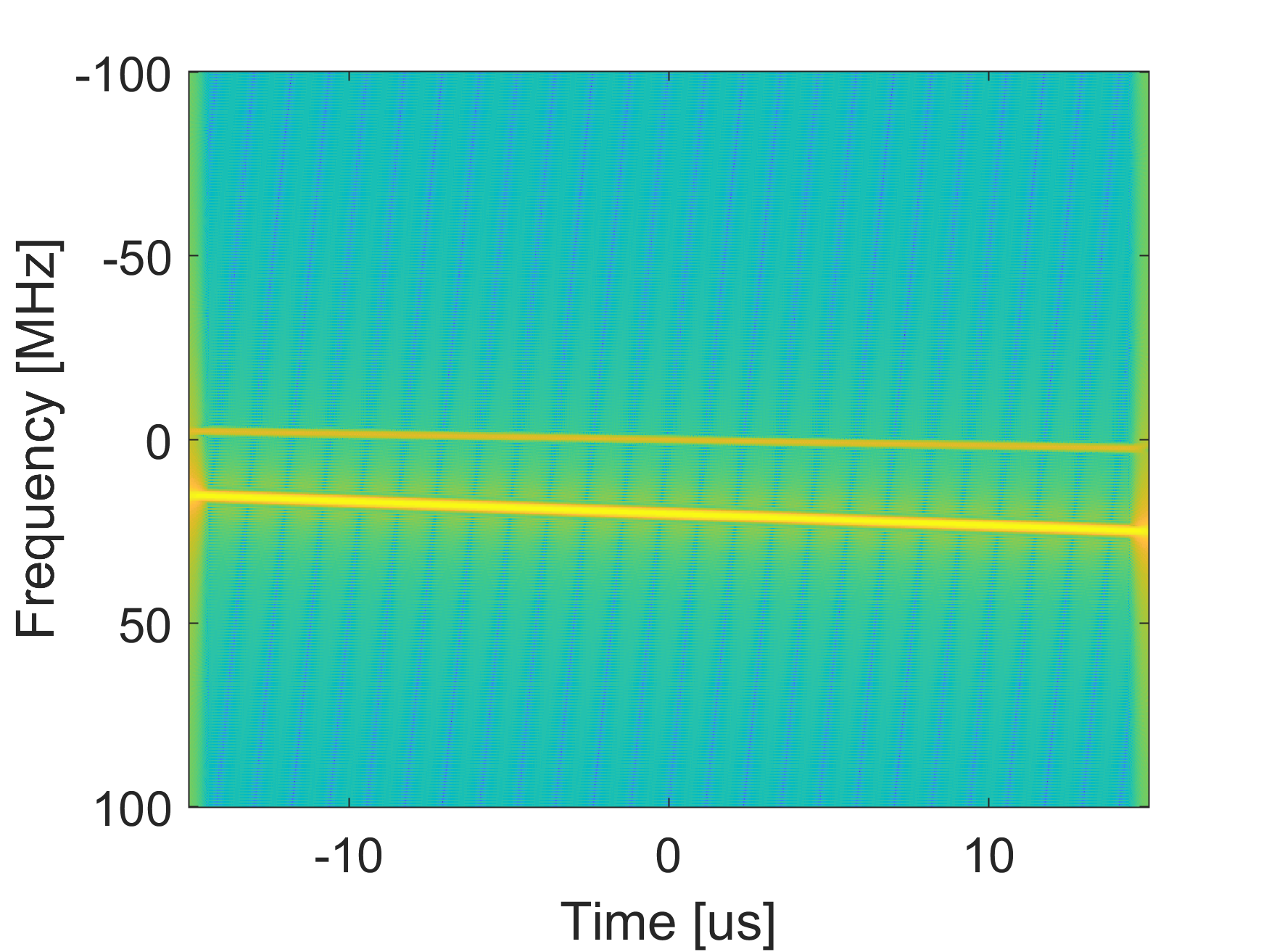}%
		\label{fig_first_case55}}\hspace{0.01in} 
	\subfloat[]{\includegraphics[width=2in]{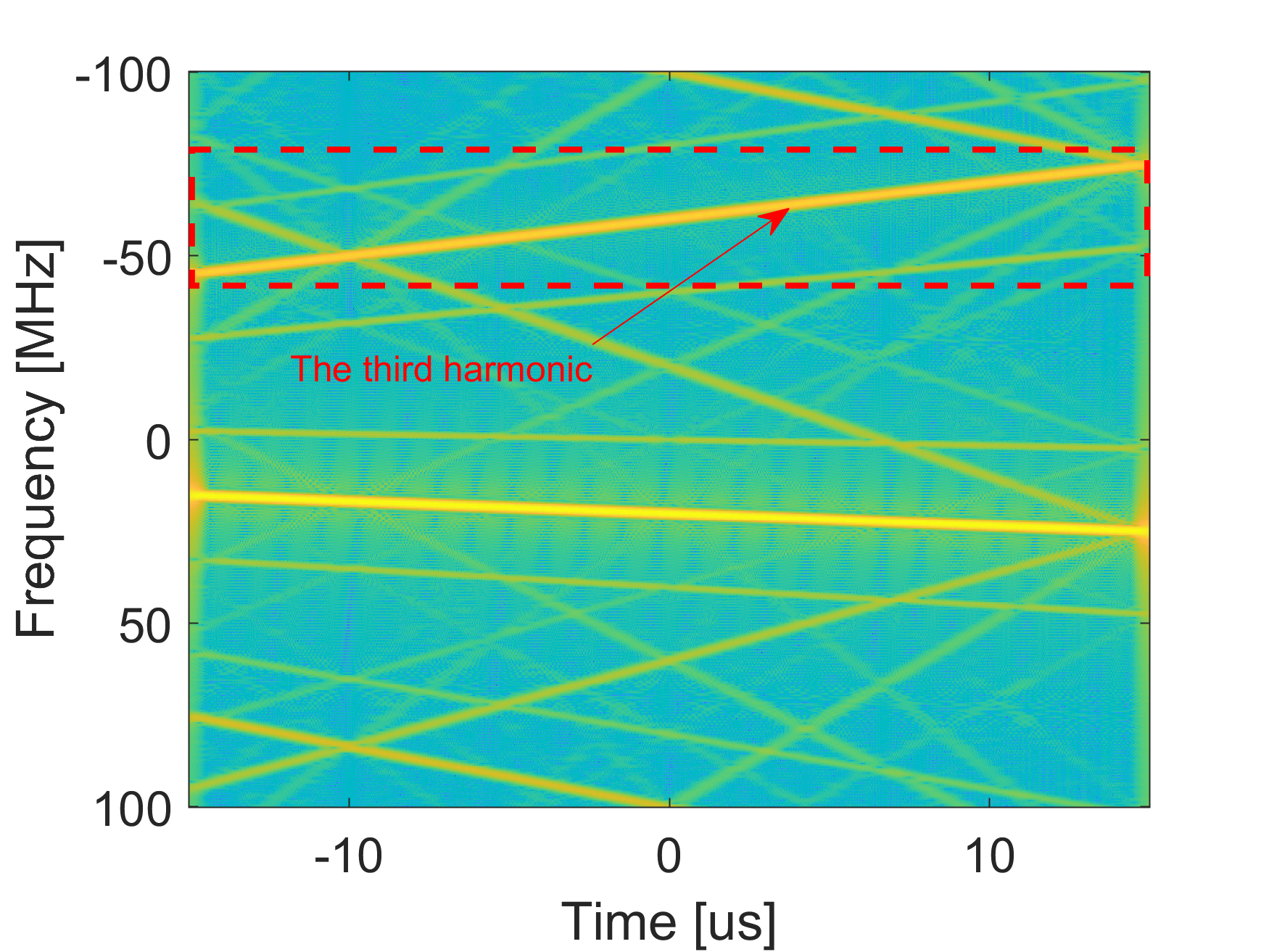}%
		\label{fig_second_case66}}\hspace{0.01in} 
	\subfloat[]{\includegraphics[width=2in]{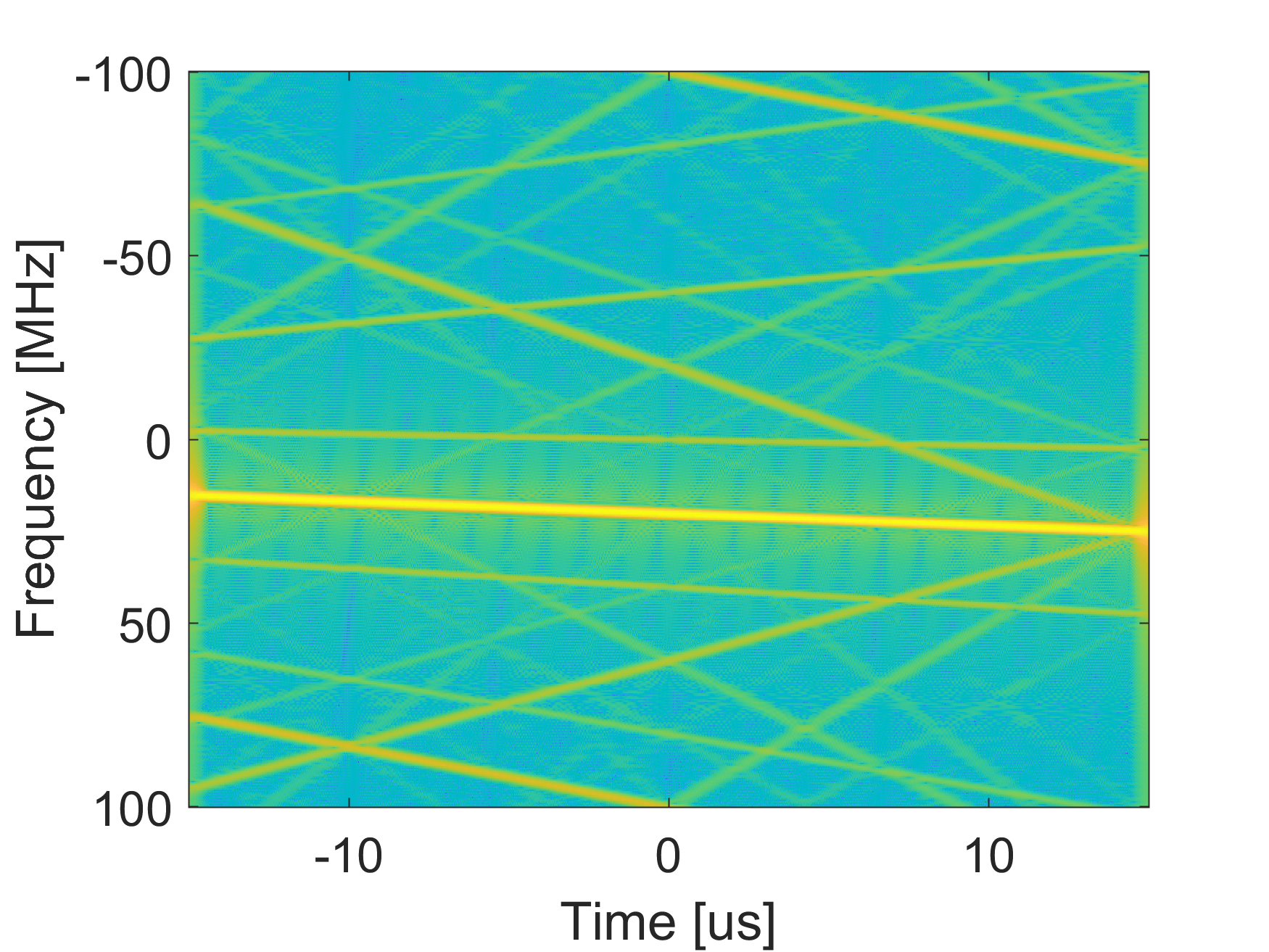}%
		\label{fig_first_case7}}

	\caption{Simulation results (a) Frequency spectrum of unsaturated signal
(b) Frequency spectrum of saturated signal
(c) Frequency spectrum after canceling the third harmonic of interference
(d) Time-frequency spectrum of unsaturated signal
(e) Time-frequency spectrum of saturated signal
(f) Time-frequency spectrum after canceling the third harmonic of interference }
	\label{fig_sim13131313}
\end{figure*}

Figs. 2(a) and 2(b) show the time-domain of the signal before and after saturation. It can be observed that the saturated signal exhibits amplitude clipping in the time domain. Figs. 3(a), 3(b), 3(d), and 3(e) present the frequency spectrum and time-frequency spectrum. Before saturation, the frequency spectrum contains interference and target echoes centered at 20 MHz and 0 MHz, respectively. After saturation, several spurious components appear in the frequency spectrum. The harmonic components generated by saturation are analyzed in detail below in conjunction with Equation (10).Observing the harmonic components marked by the red rectangular box in Fig. 3(b), the central frequency are approximately -60 MHz with a bandwidth of around 30 MHz. The corresponding component in Fig. 3(e), also marked by the red rectangular box, is observed to be LFM signal with a chirp rate approximately -3 times that of the original interference signal.

To more accurately and intuitively demonstrate the accuracy of the saturated interference analysis model, the harmonic amplitude and phase information calculated by the model can be used to reconstruct the harmonic component. By canceling the reconstructed harmonic from the saturated signal, if the residual signal no longer contains the corresponding harmonic, the accuracy of the saturation analysis model is verified.

Based on the previous analysis, we can determine that the harmonic component marked by the red rectangular box in Fig. 3(e) is the third harmonic of the interference, that is, $m=0,n=3$. Therefore, this harmonic can be expressed as follows.
\begin{equation}
\begin{split}
s_r&=\sigma[\mathrm{exp}(-\mathrm{j}(3\xi))+\mathrm{exp}(-\mathrm{j}3\xi)]\\
&=2\sigma\mathrm{exp}(-\mathrm{j}3\xi)
\end{split}
\end{equation}
where $\sigma=\frac{-\alpha_0\alpha_3A_2(-1)^2}{2\pi}=\frac{-A_2}{\pi}$

In addition, according to the previously set simulation conditions, the calculated value of $\sigma$ is -2.17. Substituting $\sigma$ into Equation (11) allows the construction of the third harmonic of the interference, and the cancellation results are shown in Figs. 3(c) and 3(f). It can be observed that after cancellation, there is no residual of the third harmonic of the interference at the corresponding positions in the frequency domain and time-frequency domain, indicating that this harmonic has been completely eliminated. This demonstrates that the saturated interference analysis model established in this paper has a certain degree of accuracy, including the accuracy of harmonic phase and harmonic amplitude.

\section{Comparative Analysis with the Saturation Model Based on the tanh Function
}
According to \cite{duan2025saturated}, by substituting Equations (3) and (4) into Equation (5), a saturation model based on the tanh function can be constructed. The first five orders of the interference higher-order harmonics are shown in Equation (13).

\begin{equation}
\begin{split}
s_{\text{out}} &= \left(b - \frac{b}{4C^{2}} + \frac{b}{12C^{4}}\right)e^{\mathrm{j}\xi} 
\quad - \left(\frac{b}{12C^{2}} - \frac{b}{24C^{4}}\right)e^{-\mathrm{j}3\xi} \\
&\quad + \frac{b}{120C^{4}}e^{\mathrm{j}5\xi}
\end{split}
\end{equation}
where $C$ denotes the saturation coefficient.

The comparative verification is still carried out using the simulation in Section 3. According to the simulation settings in Section 3, the amplitude of the third harmonic of the interference calculated using Equation (13) is -10.53. Similarly, the third harmonic of the interference is constructed and subjected to cancellation. The results after cancellation are shown in Fig. 4.
It can be seen that residual component of the third harmonic of the interference still remain after cancellation. In contrast, in the simulation results of Section 3, the third harmonic of the interference can be completely eliminated. This indicates that the Bessel function-based saturation model exhibits extremely high accuracy.

\begin{figure}[htbp]
	\centering
	\subfloat[]
	{\includegraphics[width=0.23\textwidth]{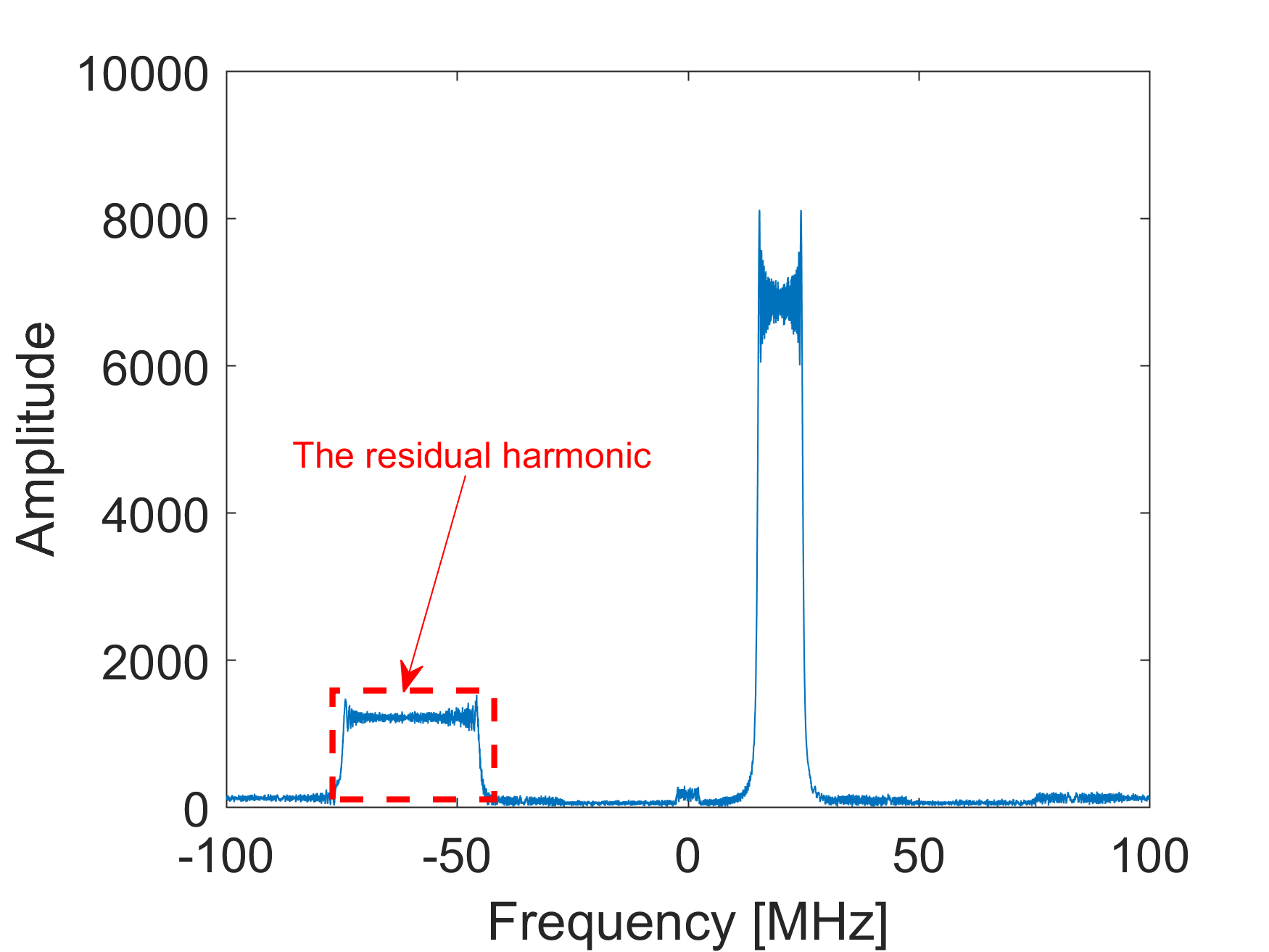}\label{figures/echo_before_sat.png}}
	\subfloat[]
	{\includegraphics[width=0.23\textwidth]{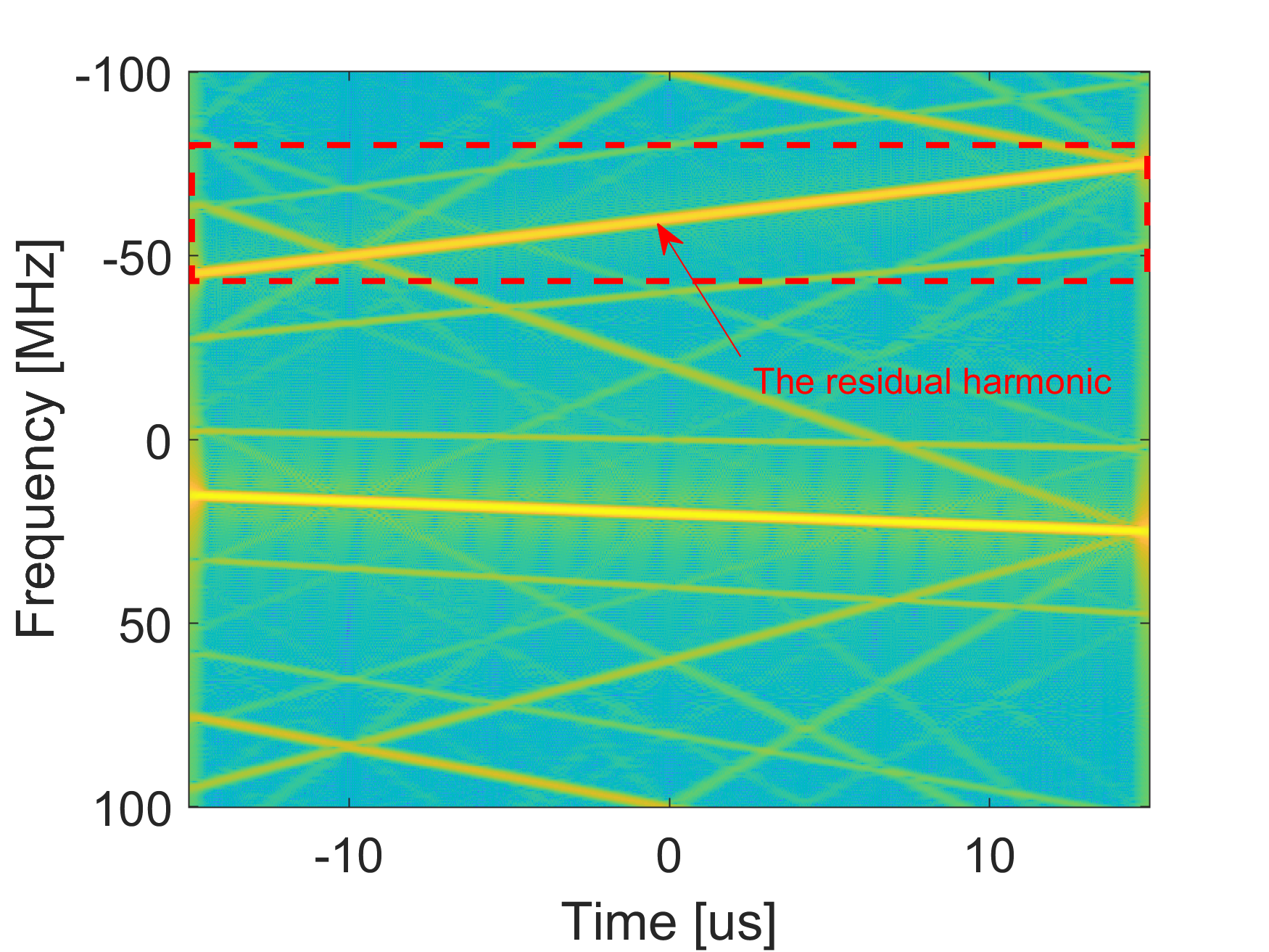}\label{figures/echo_after_sat.png}}
	\caption{Cancellation results based on the tanh function saturation model (a) Frequency spectrum after canceling the third harmonic (b) Time-frequency spectrum after canceling the third harmonic}
\end{figure}

\section{Conclusion}
For interfered saturated echoes in SAR, non-linear saturation leads to the generation of various new interference components. Existing SAR interference suppression methods have a certain degree of model mismatch, and currently there is no model that can accurately describe the SAR saturated echoes impacted by interference. Therefore, this paper proposes an accurate saturated output model based on the Bessel function. Through comparative simulations with the saturation model based on the tanh function, the accuracy of the proposed model is verified, which provides certain guidance for subsequent further analysis of the characteristics of saturated interference signals and interference suppression.

\bibliographystyle{IEEEtran}
\bibliography{main}

\vfill

\end{document}